\documentclass[prl,twocolumn,superscriptaddress]{revtex4-2}

\usepackage{amsmath,amssymb,mathrsfs}
\usepackage{graphicx}
\usepackage{colortbl}
\usepackage{braket}
\usepackage{bm}
\usepackage{amsfonts}

\begin{document}
\title{Topological Feature of Real-time Fisher Zeros}

\author{Yuchen Meng}
\affiliation {Key Laboratory of Polar Materials and Devices (MOE), School of Physics and Electronic Science, East China Normal University, Shanghai 200241, China}

\author{Yang Liu}
\altaffiliation{liuyang@stu.ecnu.edu.cn}
\affiliation{Key Laboratory of Polar Materials and Devices (MOE), School of Physics and Electronic Science, East China Normal University, Shanghai 200241, China}

\author{Erhai Zhao}
\affiliation{Department of Physics and Astronomy, George Mason University, Fairfax, Virginia 22030, USA}

\author{Haiyuan Zou}
\altaffiliation{hyzou@phy.ecnu.edu.cn}
\affiliation{Key Laboratory of Polar Materials and Devices (MOE), School of Physics and Electronic Science, East China Normal University, Shanghai 200241, China}

\begin{abstract}
There are numerous methods to characterize topology and its boundary zero modes, yet their statistical mechanical properties have not received as much attention as other approaches. Here, we investigate the Fisher zeros and thermofield dynamics of topological models, revealing that boundary zero modes can be described by an overlooked real-time Fisher zero pairing effect. This effect is validated in the Su-Schrieffer-Heeger model and the Kitaev chain model, with the latter exhibiting a Fisher zero braiding picture. Topological zero modes exhibit robustness even when non-Hermiticity is introduced into the system and display characteristics of imaginary-time crystals when the energy eigenvalues are complex.
We further examine the real-time Fisher zeros of the one-dimensional transverse field Ising model, which maps to the Kitaev chain. We present a fractal picture of the Fisher zeros, illustrating how interactions eliminate topology.
The mechanism of zero-pairing provides a natural statistical mechanical approach to understanding the connection between topology and many-body physics.
\end{abstract}

\maketitle

{\it Introduction.}
Boundary zero modes typically signify topologically non-trivial properties and are protected by symmetries~\cite{Qi2011RMP,TP1review2008,TP2review2016,TP3review2010,TP4review2016}, which result in topological invariants, such as the winding number in the one-dimensional Su-Schrieffer-Heeger (SSH) model~\cite{SSH1980PRB} and the $Z_2$ topological number in the Kitaev chain~\cite{Kitaev2001PU}. In the latter, the boundary zero modes are Majorana fermions with non-Abelian statistics, which exhibit higher robustness, making them a feasible tool for topological quantum computing~\cite{Kitaev2003anyons,Nayak2008RMP}. Apart from topological invariants, whether topological zero modes have statistical mechanical or thermofield dynamics (TFD) effects is an important issue that has been overlooked. Additionally, a direct connection can be established between the Kitaev chain and the one-dimensional transverse field Ising model (1DTFIM) through the Jordan-Wigner transformation, and the topological phase of the fermionic system corresponds to the symmetry-broken ordered phase in the spin system.
Although there exists the correspondence between the Fock space of local fermionic modes (LFMs) and the many-body Hilbert space~\cite{Bravyi2002Kitaev}, and the disappearance of topology in the many-body ordered phase can be qualitatively described using quasiparticle picture, the deep correspondence between topology and the ordered phase remains puzzling. Using statistical mechanics or TFD may provide an alternative approach to quantitatively understanding the correspondence between the two.

\begin{figure}[t]
    \includegraphics[width=0.4\textwidth]{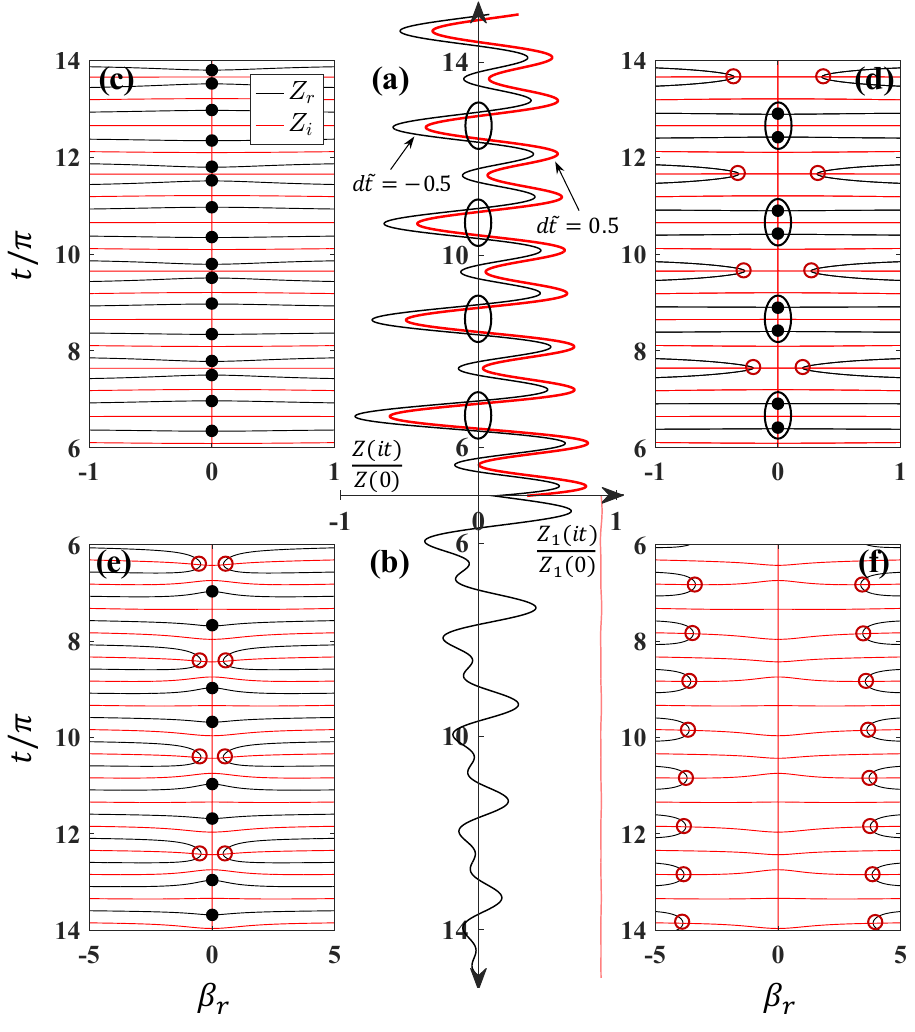}   
    \caption{Correspondence between real-time Fisher zero pairing and topological zero modes (illustrated by the SSH model with $L=20$). (a) The oscillation of $Z(it)$ as a function of $t$ for the trivial case ($d\tilde{t}=-0.5$, black line) and the topological case ($d\tilde{t}=0.5$, red line). (b) The oscillation of $Z_1$ for the two cases, where the data for $d\tilde{t}=-0.5$ has been multiplied by a factor of ten. (c,d) The Fisher zeros of $Z$ (black dots on the $t$-axis, red dots off the $t$-axis) for the two cases, respectively. The Fisher zero pairings in (a) and (d) are marked by ellipses. (e,f) The Fisher zeros of $Z_1$ for the two cases, respectively. For $d\tilde{t}=0.5$, no Fisher zeros appear on the $t$-axis.}
\label{fig:fig1}
\end{figure}

In this work, we point out that Fisher zeros, originally introduced as an extension of the Lee-Yang complexified partition function to understand thermal phase transitions~\cite{LeeYang1,LeeYang2,Fisher1965statistical}, can effectively characterize topological zero modes. In recent years, complex partition functions and Fisher zeros have regained attention. Inspired by the Lee-Yang-Fisher phase transition theory, the Fisher zeros of the boundary partition function $Z_i(\beta=\beta_r+it)=\langle\psi_i|e^{-\beta H}|\psi_i\rangle$, for a given initial state $|\psi_i\rangle$, can be used to characterize dynamical quantum phase transitions~\cite{Heyl2013PRL,Heyl2018,Zvyagin2016}. Recently, we have found that in strongly correlated many-body spin systems, the Fisher zeros of the total partition function $Z=\sum_iZ_i$ provide a quantitative tool to describe both thermal and quantum fluctuations~\cite{Liu2024CPL}. 
\begin{widetext}

\begin{figure}[t]
    \includegraphics[width=1\textwidth]{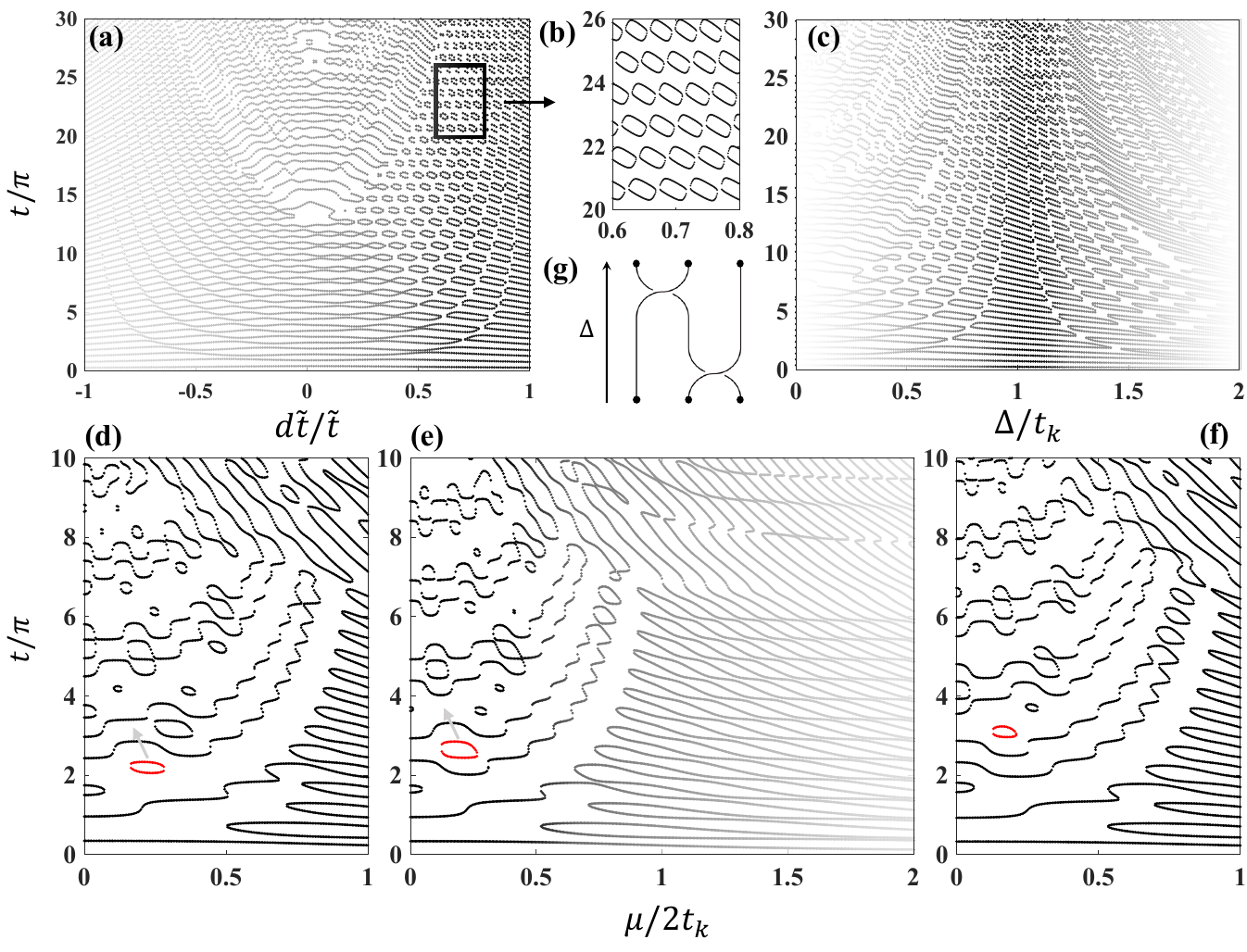}
    \caption{Real-time Fisher zeros form loop structure in the $t$-parameter space. (a) Real-time Fisher zeros in the $t-d\tilde{t}$ plane for the SSH model with $L=20$. The zeros of the system with larger size $L=40$ in the highlighted region is displayed in (b). (c) Real-time Fisher zeros in the $t-\Delta$ plane for the Kitaev chain at $\mu=0$ with $L=20$. (d-f) Real-time Fisher zeros in the $t-\mu$ plane for the Kitaev chain with $L=20$ and different values of $\Delta=0.45$, 0.5, and 0.55, respectively. One of the loops is marked in red, and the arrows indicate the moving trend of this loop as $\Delta$ increases. In (d) and (f), the region where $\mu>2t_k$ is not shown, as its Fisher zero remains largely unchanged compared to (e). (g) The motion of loop structure exhibiting braiding characteristics of Fisher zeros as $\Delta$ varies (schematic). }
    \label{fig:fig2}
\end{figure}

\end{widetext}
Many-body interactions lead to intricate TFD, manifesting in the transition of Fisher zeros from discrete points to continuous zero lines or loop structures in the entire complex $\beta$ plane as the system size increases toward the thermodynamic limit~\cite{Liu2023CPL,Liu2024PRR}. These structures effectively characterize quantum critical behavior, low-energy excitations~\cite{Liu2023CPL,Liu2024PRR}, and provide insights into the mechanisms that break the eigenstate thermalization hypothesis~\cite{Meng2025arxiv}. For one-dimensional (1D) weakly interacting fermionic models, the system's energy levels are simply obtained from the Fock space of LFMs, and $Z$ does not exhibit the complex TFD seen in many-body systems. Consequently, the corresponding Fisher zeros do not form intricate structures across the entire complex $\beta$-plane but instead cluster near the $t$-axis. Moreover, topological zero modes in fermionic models can already manifest at finite system sizes, which motivates us to explore the relationship between real-time (or imaginary inverse temperature~\cite{LiuYinChen2024}) Fisher zeros [solution for $t$ where $Z(it)=0$] and topological zero modes in finite systems.

As a prelude, we propose the following conjecture regarding the correspondence between topological zero modes and real-time Fisher zeros: {\it the zero modes in a topological phase can lead to the pairing of real-time Fisher zeros.} We first provide a discussion of Fisher zero pairing using a simple example. Consider a free fermionic system with $L$ sites under open boundary conditions (OBC), where the energy levels are symmetric, consisting of $L/2$ levels at $E_0$ and $L/2$ levels at $-E_0$. The partition function is given by $Z=L\cosh\beta E_0$. When $\beta=it$, $Z$ exhibits symmetric oscillations between positive and negative values, and the corresponding Fisher zeros are evenly spaced along the $t$-axis. In contrast, when topology is introduced into the system, one pair of energy levels is set to zero, modifying the partition function to $Z=(L-2)\cosh\beta E_0+2$. The contribution of the zero modes increases the overall values of $Z(it)$, breaking its symmetry and leading to the pairing of Fisher zeros along the $t$-axis. 
In real physical systems, the energy spectrum leads to richer oscillatory behavior of $Z(it)$, and in the absence of topology, the Fisher zeros are no longer evenly spaced. By adiabatically tuning a system parameter $g$, the actual energy levels evolve gradually, causing real-time Fisher zeros to shift along the $t$-axis. In the topological phase, the pairing effect of Fisher zeros may result in the annihilation of two closely spaced zeros, thereby forming characteristic loop structures in the $t-g$ plane.

We verify the above conjecture by studying two prototypical models with topological zero modes: the 1D SSH model and the Kitaev chain. In both models, real-time Fisher zeros exhibit a pairing effect in the topological phase. By tuning model parameters, we observe that in the time-parameter plane of the topological phase, the SSH model displays regular loop structures, while the Kitaev chain further exhibits braiding behavior of real-time Fisher zeros. By considering the non-Hermitian SSH model~\cite{WangZ2018PRL,Chen2014pra}, we not only confirm the robustness of real-time Fisher zero pairing but also uncover the phenomenon of imaginary time crystal behavior~\cite{Cai2020CPL,Smith2022iTC} when the system possesses complex energy eigenvalues. Finally, we find a fractal scaling behavior of real-time Fisher zero pairing in the 1DTFIM as the system size increases. This reveals the mechanism by which topological zero modes vanish under strong interaction in the 1DTFIM.

{\it The SSH model and Kitaev chain.} 
The SSH model, whose Hamiltonian is given by
\begin{equation}
{H} = -\tilde{t}_1\sum_{i}(\hat{c}^\dagger_{Ai}\hat{c}_{Bi}+{\rm h.c})\\
-\tilde{t}_2\sum_{i}(\hat{c}^\dagger_{Bi}\hat{c}_{Ai+1}+{\rm h.c}),
 \label{eq:HamSSH}
\end{equation}
with $\tilde{t}_1=\tilde{t}-d\tilde{t}$ and $\tilde{t}_2=\tilde{t}+d\tilde{t}$ (and set $\tilde{t}$ to 1), serves as a fundamental model for studying topological phase transitions. In this model, $\tilde{t}_{1,2}$ represents the hopping amplitudes between sublattices $A$s and $B$s. Under OBC and in the thermodynamic limit, $d\tilde{t}=0$ marks the topological phase transition point, separating the topologically trivial phase ($d\tilde{t}<0$) from the topological insulator phase ($d\tilde{t}>0$). 

\begin{figure}[t]
    \includegraphics[width=0.49\textwidth]{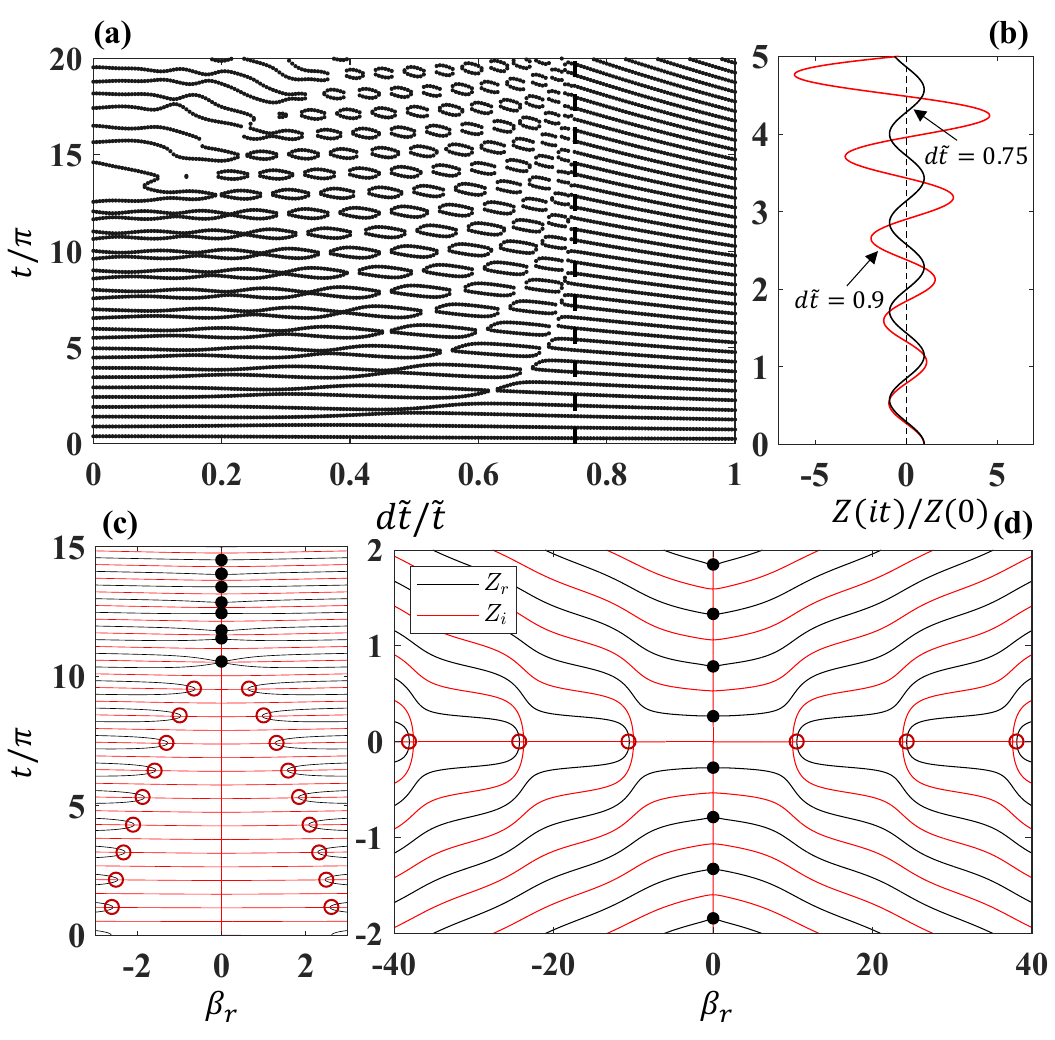}%
    \caption{Pairing of Fisher zeros in the non-Hermitian SSH model (example of $L=20$ and $\gamma=0.5$). (a) The real-time Fisher zeros also exhibit loop structures in the $t-d\tilde{t}$ plane. The region to the right of the dashed line ($d\tilde{t}=0.75$) corresponds to cases where the energy eigenvalues become complex, causing the pairing of Fisher zeros to be pinned. (b) For $d\tilde{t}=0.75$, $Z(it)$ oscillates along the $t$-axis (black line), whereas for $d\tilde{t}=0.9$, an amplification effect appears in the oscillations (red line). (c) When $d\tilde{t}=0.9$, the Fisher zeros of $Z_1$ tend to approach the $t$-axis at long time (on-axis and off-axis zeros are represented by red and black dots, respectively). (d) For $d\tilde{t}=0.9$, the Fisher zeros of $Z$ exhibit periodicity along the $\beta_r$-axis.}
    \label{fig:fig3}
\end{figure}

Taking a finite system ($L=20$) as an example, the real-time Fisher zeros exhibit a tendency to pair up when $d\tilde{t}>0$, in contrast to the case of $d\tilde{t}<0$ [Fig.~\ref{fig:fig1}(a)]. A more effective way to reveal this distinction is by solving for the Fisher zeros of $Z$ in the entire complex $\beta$ plane. In the topologically trivial phase, most zeros are densely distributed along the $t$-axis [Fig.~\ref{fig:fig1}(c)]. In contrast, in the topological phase, additional off-axis zeros appear, effectively separating the on-axis Fisher zeros into pairs [Fig.~\ref{fig:fig1}(d)]. To further investigate this phenomenon, we consider an initial state where the fermion occupies site $i$ and solving the corresponding boundary partition function $Z_1$. In the trivial phase, $Z_1$ behaves similarly to other $Z_i$, oscillating along the $t$-axis [Fig.~\ref{fig:fig1}(b)], and the corresponding Fisher zeros in the complex $\beta$ plane are primarily concentrated near the $t$-axis [Fig.~\ref{fig:fig1}(e)]. In the topological phase, however, $Z_1$ closely approximates the zero mode state and remains relatively stable with respect to $t$ [Fig.~\ref{fig:fig1}(f)], resulting in the absence of Fisher zeros on the $t$-axis in the complex $\beta$ plane. Since $Z$ is the sum over all $Z_i$, the distinct behavior of $Z_1$ in the topological phase, which differs from all other $Z_i$, plays a crucial role in the pairing effect of real-time Fisher zeros.

By adiabatically tuning $d\tilde{t}$, the real-time Fisher zeros in the topologically trivial and nontrivial phases exhibit distinctly different patterns [Fig.~\ref{fig:fig2}(a)]. When $d\tilde{t}<0$, the real-time Fisher zeros evolve relatively continuously in the $t-d\tilde{t}$ plane as $d\tilde{t}$ varies. In contrast, for $d\tilde{t}>0$, the Fisher zeros form distinct regular loop structures, providing a more intuitive illustration of the pairing effect of Fisher zeros within the topological phase. These loop structures remain stable as $L$ increases, only exhibiting a shrinking trend [Fig.~\ref{fig:fig2}(b)]. The intermediate region near $d\tilde{t}\approx 0$ shifts towards longer time (larger $t$), as $L$ increases, further enhancing the contrast between the Fisher zero structures on the two sides of the topological phase transition point.

We then consider the Kitaev chain, whose Hamiltonian is given by
\begin{equation}
{H} = -\sum_{i}(t_k\hat{c}^\dagger_{i}\hat{c}_{i+1}\\
+\Delta\hat{c}^\dagger_{i}\hat{c}^\dagger_{i+1}+{\rm h.c})-\mu\sum_i\hat{c}^\dagger_{i}\hat{c}_{i},
 \label{eq:HamKitaev}
\end{equation}
with $t_k=1$.
This system also undergoes a topological phase transition when the superconducting pairing parameter $\Delta>0$. For $\mu<2t_k$, the system is in a topologically nontrivial 1D $p$-wave superconducting phase, where Majorana zero modes exist. We calculate the real-time Fisher zeros in the $t-\Delta$ plane at $\mu=0$, and the results also exhibit loop structure [Fig.~\ref{fig:fig2}(c)]. The similarity between the Fisher zeros of the Kitaev model in the $t-\Delta$ plane for $\Delta\lesssim 1$ and those of the SSH model in the $t-d\tilde{t}$ plane for $d\tilde{t}\lesssim 1$ reflects the fact that, at $\mu=0$, the two models can be mapped onto each other~\cite{Wakatsuki2014PRB,LiSong2018PRB}.  

We further consider the case of $\mu >0$, $\Delta>0$ and calculate the real-time Fisher zeros in the $t-\mu$ plane for different $\Delta$. In the region $\mu>2t_k$, as 
$\mu$ varies, the zeros exhibit continuity, corresponding to the topologically trivial phase [Fig.~\ref{fig:fig2}(e)]. In contrast, in the topological phase where $\mu<2t_k$, the pairing of Fisher zeros results in a loop structure that is more complex than the one observed in the SSH model. By adjusting $\Delta$, we find that in the topological region, there is a crossover behavior between zeros loops and lines [Fig.~\ref{fig:fig2}(d-f)]. This behavior can be visually represented as the braiding of real-time Fisher zeros when $\Delta$ is tuned [Fig.~\ref{fig:fig2}(g)], thus demonstrating that Fisher zeros can provide new insights into the braiding behavior of Majorana fermions when adjusting the superconducting gap.

\begin{figure}[t]
    \includegraphics[width=0.49\textwidth]{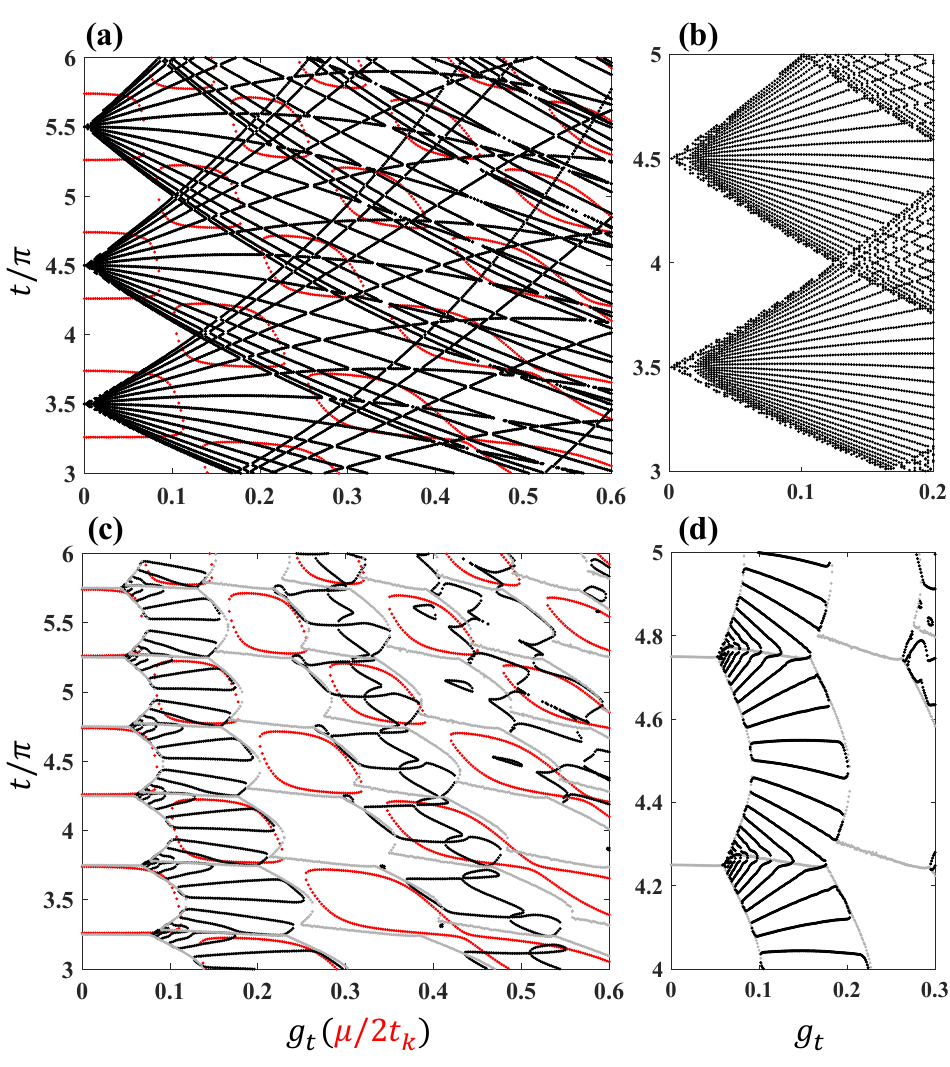}%
    \caption{Real-time Fisher zeros of the 1DTFIM (Kitaev chain with $t_k=\Delta$) in $t-g_t$ ($t-\mu$) plane under different boundary conditions. (a) Fisher zeros with OBC for $L=16$ (black line). Zeros of the Kitaev chain with the same system size are shown in red. (b) The Fisher zeros in a partial region in (a) for $L=32$. (c) Fisher zeros  with PBC for $L=16$ (black), alongside the Fisher zeros in the thermodynamic limit (gray) and those of the Kitaev chain (red). (d) The Fisher zeros in a partial region in (c) for $L=32$.}
    \label{fig:fig4}
\end{figure}

{~\it Non-Hermitian SSH model. }
We also consider the effect of non-Hermitian terms on real-time Fisher zero pairing using the non-Hermitian SSH model as an example. We add a non-Hermitian term to the first term in Eq.~(\ref{eq:HamSSH}), making the hopping from B to A become $\tilde{t}_1+\gamma/2$, and the hopping from A to B become $\tilde{t}_1-\gamma/2$. By introducing the generalized Brillouin zone and the non-Bloch winding number, the topology of the non-Hermitian SSH model can be well understood~\cite{WangZ2018PRL}. The research on non-Hermitian and complex partition functions, as well as Fisher zeros, is still in its early stages~\cite{Basu2022PRR,Xie2024arxiv,Liu2024CPL}. Here, we calculate the Fisher zeros of the system after introducing $\gamma$ and find that the loop structure in the $t-d\tilde{t}$ plane caused by the zero pairing effect still exists [Fig.~\ref{fig:fig3}(a)], demonstrating the effectiveness of real-time Fisher zeros pairing in characterizing topology. In the region $d\tilde{t}<1-\gamma/2$, the system's energy eigenvalues remain real, indicating the existence of an effective hermitian Hamiltonian, related by a similarity transformation, to describe the non-Hermitian system. In this region, the loop structures in the $t-d\tilde{t}$ plane are similar to the ones observed in the SSH model. For the cases when the energy eigenvalues become complex ($d\tilde{t}>1-\gamma/2$), we find that the Fisher zero pairing in the $t-d\tilde{t}$ plane exhibits a freezing effect. In this case, the system's energy levels experience decay and gain, which strengthens the oscillatory behavior of the partition function $Z$ [Fig.~\ref{fig:fig3}(b)], preventing the real-time Fisher zeros from undergoing drastic changes due to $d\tilde{t}$ tuning. By calculating $Z_1$ for the zero mode state, we observe that long-time zeros tend to approach the $t$-axis [Fig.~\ref{fig:fig3}(c)]. It is worth noting that, unlike in the real energy region, $Z$ also exhibits periodic Fisher zeros on the $\beta_r$ axis [Fig.~\ref{fig:fig3}(d)]. This result is consistent with the previous divergent specific heat result~\cite{Smith2022iTC} and suggests the possibility of realizing an imaginary-time crystal~\cite{Cai2020CPL} using a non-Hermitian system.

{~\it 1DTFIM. }
Through the Jordan-Wigner transformation, the Kitaev chain at $t_k=\Delta$ can be mapped to the 1DTFIM, with the Hamiltonian given by
\begin{equation}
    {H} = -\sum_{i}\left(\sigma_i^z\sigma_{i+1}^z + g_t\sigma_i^x \right),
    \label{eq:HamTFIM}
\end{equation}
where $g_t$ is the strength of the transverse field.
Although the phase transitions in these two models correspond to each other, the 1DTFIM no longer exhibits a topological phase. Under OBC, the zero modes disappear entirely, and the topological order in the Kitaev chain is replaced by an ordered spin phase. When transitioning from the single-particle fermionic picture to the spin model, the system's Hilbert space expands, and the sector originally described by the single-particle Hamiltonian becomes coupled to other parts of the Hilbert space. From the quasiparticle perspective, the previously robust boundary states can propagate via domain walls to other states, thereby destroying the topological nature. 
Using the exact solution of the energy spectrum of the 1DTFIM under OBC~\cite{Lieb1961AOP}, we calculate the partition function and reveal that the real-time Fisher zero loop structures in the $t$-$g_t$ plane completely vanish [Fig.~\ref{fig:fig4}(a,b)]. 

For the 1DTFIM with periodic boundary conditions (PBC), accidental zero-energy modes may still exist, but they are fundamentally different from topological zero modes. To illustrate this, we further compute the real-time Fisher zeros in finite systems. The results show that in the $t$-$g_t$ plane, the Fisher zeros form a more intricate structure [Fig.~\ref{fig:fig4}(c,d)]. However, as the system size increases, the loops of Fisher zeros become denser and gradually fill the entire envelope region formed by Fisher zeros in the thermodynamic limit. This envelope region exhibits an overall elongated structure, which significantly differs from the characteristic behavior of fermionic models, where the Fisher zero structure in the topological phase remains unchanged as the system size increases.

{~\it Conclusion. } 
In summary, our study establishes a direct correspondence between real-time Fisher zero pairing and topological zero modes. As the system parameter $g$ varies, Fisher zeros form loops structure in the $\beta_i$-$g$ plane. By investigating the SSH model and Kitaev chain—two prototypical models with topological zero modes—we confirm this correspondence and reveal a braiding behavior of Fisher zeros associated with Majorana fermions in the Kitaev chain. Extending the analysis to the non-Hermitian SSH model, we demonstrate the robustness of real-time Fisher zero pairing and uncover the emergence of an imaginary time crystal when complex energy eigenvalues appear. Additionally, in the 1DTFIM, we observe fractal scaling of Fisher zero pairing with system size, shedding light on the mechanism by which topological zero modes vanish under strong interactions. Our findings offer new perspectives for exploring richer topological behaviors in high-dimensional~\cite{Haldane1088prl,QiWuZ2006PRB,BHZ2006} and many-body systems~\cite{KitaevPreskill2006TEE,LevinWen2006prl,LiHaldane2008prl,SPT2011prb1} through Fisher zero configurations.

{~\it Acknowledgments.} We thank Shijie Hu, Gaoyong Sun and Liping Yang for helpful discussions. H.Z. expresses special thanks to Prof. Olexei Motrunich, who mentioned our work on Fisher zeros in his latest preprint on scars. This seemingly ordinary event may reflect the amazing entanglement feature of science and education across space and time, because 13 years ago, H.Z. stumbled upon Prof. Motrunich's beautiful handwritten 127c notes online to learn advanced statistical physics and benefited greatly from them. This work is supported by the National Natural Science Foundation of China (Grant No. 12274126). E.Z. acknowledges the support from NSF Grant PHY-206419, and AFOSR Grant FA9550-23-1-0598. 


\begin{thebibliography}{37}%
\makeatletter
\providecommand \@ifxundefined [1]{%
 \@ifx{#1\undefined}
}%
\providecommand \@ifnum [1]{%
 \ifnum #1\expandafter \@firstoftwo
 \else \expandafter \@secondoftwo
 \fi
}%
\providecommand \@ifx [1]{%
 \ifx #1\expandafter \@firstoftwo
 \else \expandafter \@secondoftwo
 \fi
}%
\providecommand \natexlab [1]{#1}%
\providecommand \enquote  [1]{``#1''}%
\providecommand \bibnamefont  [1]{#1}%
\providecommand \bibfnamefont [1]{#1}%
\providecommand \citenamefont [1]{#1}%
\providecommand \href@noop [0]{\@secondoftwo}%
\providecommand \href [0]{\begingroup \@sanitize@url \@href}%
\providecommand \@href[1]{\@@startlink{#1}\@@href}%
\providecommand \@@href[1]{\endgroup#1\@@endlink}%
\providecommand \@sanitize@url [0]{\catcode `\\12\catcode `\$12\catcode
  `\&12\catcode `\#12\catcode `\^12\catcode `\_12\catcode `\%12\relax}%
\providecommand \@@startlink[1]{}%
\providecommand \@@endlink[0]{}%
\providecommand \url  [0]{\begingroup\@sanitize@url \@url }%
\providecommand \@url [1]{\endgroup\@href {#1}{\urlprefix }}%
\providecommand \urlprefix  [0]{URL }%
\providecommand \Eprint [0]{\href }%
\providecommand \doibase [0]{https://doi.org/}%
\providecommand \selectlanguage [0]{\@gobble}%
\providecommand \bibinfo  [0]{\@secondoftwo}%
\providecommand \bibfield  [0]{\@secondoftwo}%
\providecommand \translation [1]{[#1]}%
\providecommand \BibitemOpen [0]{}%
\providecommand \bibitemStop [0]{}%
\providecommand \bibitemNoStop [0]{.\EOS\space}%
\providecommand \EOS [0]{\spacefactor3000\relax}%
\providecommand \BibitemShut  [1]{\csname bibitem#1\endcsname}%
\let\auto@bib@innerbib\@empty
\bibitem [{\citenamefont {Qi}\ and\ \citenamefont {Zhang}(2011)}]{Qi2011RMP}%
  \BibitemOpen
  \bibfield  {author} {\bibinfo {author} {\bibfnamefont {X.-L.}\ \bibnamefont
  {Qi}}\ and\ \bibinfo {author} {\bibfnamefont {S.-C.}\ \bibnamefont {Zhang}},\
  }\bibfield  {title} {\bibinfo {title} {Topological insulators and
  superconductors},\ }\href {https://doi.org/10.1103/RevModPhys.83.1057}
  {\bibfield  {journal} {\bibinfo  {journal} {Rev. Mod. Phys.}\ }\textbf
  {\bibinfo {volume} {83}},\ \bibinfo {pages} {1057} (\bibinfo {year}
  {2011})}\BibitemShut {NoStop}%
\bibitem [{\citenamefont {Nayak}\ \emph
  {et~al.}(2008{\natexlab{a}})\citenamefont {Nayak}, \citenamefont {Simon},
  \citenamefont {Stern}, \citenamefont {Freedman},\ and\ \citenamefont
  {Das~Sarma}}]{TP1review2008}%
  \BibitemOpen
  \bibfield  {author} {\bibinfo {author} {\bibfnamefont {C.}~\bibnamefont
  {Nayak}}, \bibinfo {author} {\bibfnamefont {S.~H.}\ \bibnamefont {Simon}},
  \bibinfo {author} {\bibfnamefont {A.}~\bibnamefont {Stern}}, \bibinfo
  {author} {\bibfnamefont {M.}~\bibnamefont {Freedman}},\ and\ \bibinfo
  {author} {\bibfnamefont {S.}~\bibnamefont {Das~Sarma}},\ }\bibfield  {title}
  {\bibinfo {title} {Non-abelian anyons and topological quantum computation},\
  }\href {https://doi.org/10.1103/RevModPhys.80.1083} {\bibfield  {journal}
  {\bibinfo  {journal} {Rev. Mod. Phys.}\ }\textbf {\bibinfo {volume} {80}},\
  \bibinfo {pages} {1083} (\bibinfo {year} {2008}{\natexlab{a}})}\BibitemShut
  {NoStop}%
\bibitem [{\citenamefont {Chiu}\ \emph {et~al.}(2016)\citenamefont {Chiu},
  \citenamefont {Teo}, \citenamefont {Schnyder},\ and\ \citenamefont
  {Ryu}}]{TP2review2016}%
  \BibitemOpen
  \bibfield  {author} {\bibinfo {author} {\bibfnamefont {C.-K.}\ \bibnamefont
  {Chiu}}, \bibinfo {author} {\bibfnamefont {J.~C.~Y.}\ \bibnamefont {Teo}},
  \bibinfo {author} {\bibfnamefont {A.~P.}\ \bibnamefont {Schnyder}},\ and\
  \bibinfo {author} {\bibfnamefont {S.}~\bibnamefont {Ryu}},\ }\bibfield
  {title} {\bibinfo {title} {Classification of topological quantum matter with
  symmetries},\ }\href {https://doi.org/10.1103/RevModPhys.88.035005}
  {\bibfield  {journal} {\bibinfo  {journal} {Rev. Mod. Phys.}\ }\textbf
  {\bibinfo {volume} {88}},\ \bibinfo {pages} {035005} (\bibinfo {year}
  {2016})}\BibitemShut {NoStop}%
\bibitem [{\citenamefont {Hasan}\ and\ \citenamefont
  {Kane}(2010)}]{TP3review2010}%
  \BibitemOpen
  \bibfield  {author} {\bibinfo {author} {\bibfnamefont {M.~Z.}\ \bibnamefont
  {Hasan}}\ and\ \bibinfo {author} {\bibfnamefont {C.~L.}\ \bibnamefont
  {Kane}},\ }\bibfield  {title} {\bibinfo {title} {Colloquium: Topological
  insulators},\ }\href {https://doi.org/10.1103/RevModPhys.82.3045} {\bibfield
  {journal} {\bibinfo  {journal} {Rev. Mod. Phys.}\ }\textbf {\bibinfo {volume}
  {82}},\ \bibinfo {pages} {3045} (\bibinfo {year} {2010})}\BibitemShut
  {NoStop}%
\bibitem [{\citenamefont {Bansil}\ \emph {et~al.}(2016)\citenamefont {Bansil},
  \citenamefont {Lin},\ and\ \citenamefont {Das}}]{TP4review2016}%
  \BibitemOpen
  \bibfield  {author} {\bibinfo {author} {\bibfnamefont {A.}~\bibnamefont
  {Bansil}}, \bibinfo {author} {\bibfnamefont {H.}~\bibnamefont {Lin}},\ and\
  \bibinfo {author} {\bibfnamefont {T.}~\bibnamefont {Das}},\ }\bibfield
  {title} {\bibinfo {title} {Colloquium: Topological band theory},\ }\bibfield
  {journal} {\bibinfo  {journal} {Reviews of Modern Physics}\ }\textbf
  {\bibinfo {volume} {88}},\ \href
  {https://doi.org/10.1103/revmodphys.88.021004} {10.1103/revmodphys.88.021004}
  (\bibinfo {year} {2016})\BibitemShut {NoStop}%
\bibitem [{\citenamefont {Su}\ \emph {et~al.}(1980)\citenamefont {Su},
  \citenamefont {Schrieffer},\ and\ \citenamefont {Heeger}}]{SSH1980PRB}%
  \BibitemOpen
  \bibfield  {author} {\bibinfo {author} {\bibfnamefont {W.~P.}\ \bibnamefont
  {Su}}, \bibinfo {author} {\bibfnamefont {J.~R.}\ \bibnamefont {Schrieffer}},\
  and\ \bibinfo {author} {\bibfnamefont {A.~J.}\ \bibnamefont {Heeger}},\
  }\bibfield  {title} {\bibinfo {title} {Soliton excitations in
  polyacetylene},\ }\href {https://doi.org/10.1103/PhysRevB.22.2099} {\bibfield
   {journal} {\bibinfo  {journal} {Phys. Rev. B}\ }\textbf {\bibinfo {volume}
  {22}},\ \bibinfo {pages} {2099} (\bibinfo {year} {1980})}\BibitemShut
  {NoStop}%
\bibitem [{\citenamefont {Kitaev}(2001)}]{Kitaev2001PU}%
  \BibitemOpen
  \bibfield  {author} {\bibinfo {author} {\bibfnamefont {A.~Y.}\ \bibnamefont
  {Kitaev}},\ }\bibfield  {title} {\bibinfo {title} {Unpaired majorana fermions
  in quantum wires},\ }\href {https://doi.org/10.1070/1063-7869/44/10S/S29}
  {\bibfield  {journal} {\bibinfo  {journal} {Physics-Uspekhi}\ }\textbf
  {\bibinfo {volume} {44}},\ \bibinfo {pages} {131} (\bibinfo {year}
  {2001})}\BibitemShut {NoStop}%
\bibitem [{\citenamefont {Kitaev}(2003)}]{Kitaev2003anyons}%
  \BibitemOpen
  \bibfield  {author} {\bibinfo {author} {\bibfnamefont {A.}~\bibnamefont
  {Kitaev}},\ }\bibfield  {title} {\bibinfo {title} {Fault-tolerant quantum
  computation by anyons},\ }\href
  {https://doi.org/https://doi.org/10.1016/S0003-4916(02)00018-0} {\bibfield
  {journal} {\bibinfo  {journal} {Annals of Physics}\ }\textbf {\bibinfo
  {volume} {303}},\ \bibinfo {pages} {2} (\bibinfo {year} {2003})}\BibitemShut
  {NoStop}%
\bibitem [{\citenamefont {Nayak}\ \emph
  {et~al.}(2008{\natexlab{b}})\citenamefont {Nayak}, \citenamefont {Simon},
  \citenamefont {Stern}, \citenamefont {Freedman},\ and\ \citenamefont
  {Das~Sarma}}]{Nayak2008RMP}%
  \BibitemOpen
  \bibfield  {author} {\bibinfo {author} {\bibfnamefont {C.}~\bibnamefont
  {Nayak}}, \bibinfo {author} {\bibfnamefont {S.~H.}\ \bibnamefont {Simon}},
  \bibinfo {author} {\bibfnamefont {A.}~\bibnamefont {Stern}}, \bibinfo
  {author} {\bibfnamefont {M.}~\bibnamefont {Freedman}},\ and\ \bibinfo
  {author} {\bibfnamefont {S.}~\bibnamefont {Das~Sarma}},\ }\bibfield  {title}
  {\bibinfo {title} {Non-abelian anyons and topological quantum computation},\
  }\href {https://doi.org/10.1103/RevModPhys.80.1083} {\bibfield  {journal}
  {\bibinfo  {journal} {Rev. Mod. Phys.}\ }\textbf {\bibinfo {volume} {80}},\
  \bibinfo {pages} {1083} (\bibinfo {year} {2008}{\natexlab{b}})}\BibitemShut
  {NoStop}%
\bibitem [{\citenamefont {Bravyi}\ and\ \citenamefont
  {Kitaev}(2002)}]{Bravyi2002Kitaev}%
  \BibitemOpen
  \bibfield  {author} {\bibinfo {author} {\bibfnamefont {S.~B.}\ \bibnamefont
  {Bravyi}}\ and\ \bibinfo {author} {\bibfnamefont {A.~Y.}\ \bibnamefont
  {Kitaev}},\ }\bibfield  {title} {\bibinfo {title} {Fermionic quantum
  computation},\ }\href
  {https://doi.org/https://doi.org/10.1006/aphy.2002.6254} {\bibfield
  {journal} {\bibinfo  {journal} {Annals of Physics}\ }\textbf {\bibinfo
  {volume} {298}},\ \bibinfo {pages} {210} (\bibinfo {year}
  {2002})}\BibitemShut {NoStop}%
\bibitem [{\citenamefont {Yang}\ and\ \citenamefont {Lee}(1952)}]{LeeYang1}%
  \BibitemOpen
  \bibfield  {author} {\bibinfo {author} {\bibfnamefont {C.~N.}\ \bibnamefont
  {Yang}}\ and\ \bibinfo {author} {\bibfnamefont {T.~D.}\ \bibnamefont {Lee}},\
  }\bibfield  {title} {\bibinfo {title} {Statistical theory of equations of
  state and phase transitions. i. theory of condensation},\ }\href
  {https://doi.org/10.1103/PhysRev.87.404} {\bibfield  {journal} {\bibinfo
  {journal} {Phys. Rev.}\ }\textbf {\bibinfo {volume} {87}},\ \bibinfo {pages}
  {404} (\bibinfo {year} {1952})}\BibitemShut {NoStop}%
\bibitem [{\citenamefont {Lee}\ and\ \citenamefont {Yang}(1952)}]{LeeYang2}%
  \BibitemOpen
  \bibfield  {author} {\bibinfo {author} {\bibfnamefont {T.~D.}\ \bibnamefont
  {Lee}}\ and\ \bibinfo {author} {\bibfnamefont {C.~N.}\ \bibnamefont {Yang}},\
  }\bibfield  {title} {\bibinfo {title} {Statistical theory of equations of
  state and phase transitions. ii. lattice gas and ising model},\ }\href
  {https://doi.org/10.1103/PhysRev.87.410} {\bibfield  {journal} {\bibinfo
  {journal} {Phys. Rev.}\ }\textbf {\bibinfo {volume} {87}},\ \bibinfo {pages}
  {410} (\bibinfo {year} {1952})}\BibitemShut {NoStop}%
\bibitem [{\citenamefont {Fisher}\ and\ \citenamefont
  {Brittin}(1965)}]{Fisher1965statistical}%
  \BibitemOpen
  \bibfield  {author} {\bibinfo {author} {\bibfnamefont {M.}~\bibnamefont
  {Fisher}}\ and\ \bibinfo {author} {\bibfnamefont {W.}~\bibnamefont
  {Brittin}},\ }\bibfield  {title} {\bibinfo {title} {Statistical physics, weak
  interactions, field theory},\ }\href@noop {} {\bibfield  {journal} {\bibinfo
  {journal} {Lectures in Theoretical Physics (Boulder: University of Colorado
  Press) vol VIIC}\ } (\bibinfo {year} {1965})}\BibitemShut {NoStop}%
\bibitem [{\citenamefont {Heyl}\ \emph {et~al.}(2013)\citenamefont {Heyl},
  \citenamefont {Polkovnikov},\ and\ \citenamefont {Kehrein}}]{Heyl2013PRL}%
  \BibitemOpen
  \bibfield  {author} {\bibinfo {author} {\bibfnamefont {M.}~\bibnamefont
  {Heyl}}, \bibinfo {author} {\bibfnamefont {A.}~\bibnamefont {Polkovnikov}},\
  and\ \bibinfo {author} {\bibfnamefont {S.}~\bibnamefont {Kehrein}},\
  }\bibfield  {title} {\bibinfo {title} {Dynamical quantum phase transitions in
  the transverse-field ising model},\ }\href
  {https://doi.org/10.1103/PhysRevLett.110.135704} {\bibfield  {journal}
  {\bibinfo  {journal} {Phys. Rev. Lett.}\ }\textbf {\bibinfo {volume} {110}},\
  \bibinfo {pages} {135704} (\bibinfo {year} {2013})}\BibitemShut {NoStop}%
\bibitem [{\citenamefont {Heyl}(2018)}]{Heyl2018}%
  \BibitemOpen
  \bibfield  {author} {\bibinfo {author} {\bibfnamefont {M.}~\bibnamefont
  {Heyl}},\ }\bibfield  {title} {\bibinfo {title} {Dynamical quantum phase
  transitions: a review},\ }\href {https://doi.org/10.1088/1361-6633/aaaf9a}
  {\bibfield  {journal} {\bibinfo  {journal} {Reports on Progress in Physics}\
  }\textbf {\bibinfo {volume} {81}},\ \bibinfo {pages} {054001} (\bibinfo
  {year} {2018})}\BibitemShut {NoStop}%
\bibitem [{\citenamefont {Zvyagin}(2016)}]{Zvyagin2016}%
  \BibitemOpen
  \bibfield  {author} {\bibinfo {author} {\bibfnamefont {A.~A.}\ \bibnamefont
  {Zvyagin}},\ }\bibfield  {title} {\bibinfo {title} {Dynamical quantum phase
  transitions (review article)},\ }\href {https://doi.org/10.1063/1.4969869}
  {\bibfield  {journal} {\bibinfo  {journal} {Low Temperature Physics}\
  }\textbf {\bibinfo {volume} {42}},\ \bibinfo {pages} {971–994} (\bibinfo
  {year} {2016})}\BibitemShut {NoStop}%
\bibitem [{\citenamefont {Liu}\ \emph {et~al.}(2024{\natexlab{a}})\citenamefont
  {Liu}, \citenamefont {Zhao},\ and\ \citenamefont {Zou}}]{Liu2024CPL}%
  \BibitemOpen
  \bibfield  {author} {\bibinfo {author} {\bibfnamefont {Y.}~\bibnamefont
  {Liu}}, \bibinfo {author} {\bibfnamefont {E.}~\bibnamefont {Zhao}},\ and\
  \bibinfo {author} {\bibfnamefont {H.}~\bibnamefont {Zou}},\ }\bibfield
  {title} {\bibinfo {title} {From complexification to self-similarity: New
  aspects of quantum criticality},\ }\href
  {https://doi.org/10.1088/0256-307X/41/10/100501} {\bibfield  {journal}
  {\bibinfo  {journal} {Chinese Physics Letters}\ }\textbf {\bibinfo {volume}
  {41}},\ \bibinfo {pages} {100501} (\bibinfo {year}
  {2024}{\natexlab{a}})}\BibitemShut {NoStop}%
\bibitem [{\citenamefont {Liu}\ \emph {et~al.}(2023)\citenamefont {Liu},
  \citenamefont {Lv}, \citenamefont {Yang},\ and\ \citenamefont
  {Zou}}]{Liu2023CPL}%
  \BibitemOpen
  \bibfield  {author} {\bibinfo {author} {\bibfnamefont {Y.}~\bibnamefont
  {Liu}}, \bibinfo {author} {\bibfnamefont {S.}~\bibnamefont {Lv}}, \bibinfo
  {author} {\bibfnamefont {Y.}~\bibnamefont {Yang}},\ and\ \bibinfo {author}
  {\bibfnamefont {H.}~\bibnamefont {Zou}},\ }\bibfield  {title} {\bibinfo
  {title} {Signatures of quantum criticality in the complex inverse temperature
  plane},\ }\href@noop {} {\bibfield  {journal} {\bibinfo  {journal} {Chinese
  Physics Letters}\ }\textbf {\bibinfo {volume} {40}},\ \bibinfo {pages}
  {050502} (\bibinfo {year} {2023})}\BibitemShut {NoStop}%
\bibitem [{\citenamefont {Liu}\ \emph {et~al.}(2024{\natexlab{b}})\citenamefont
  {Liu}, \citenamefont {Lv}, \citenamefont {Meng}, \citenamefont {Tan},
  \citenamefont {Zhao},\ and\ \citenamefont {Zou}}]{Liu2024PRR}%
  \BibitemOpen
  \bibfield  {author} {\bibinfo {author} {\bibfnamefont {Y.}~\bibnamefont
  {Liu}}, \bibinfo {author} {\bibfnamefont {S.}~\bibnamefont {Lv}}, \bibinfo
  {author} {\bibfnamefont {Y.}~\bibnamefont {Meng}}, \bibinfo {author}
  {\bibfnamefont {Z.}~\bibnamefont {Tan}}, \bibinfo {author} {\bibfnamefont
  {E.}~\bibnamefont {Zhao}},\ and\ \bibinfo {author} {\bibfnamefont
  {H.}~\bibnamefont {Zou}},\ }\bibfield  {title} {\bibinfo {title} {Exact
  fisher zeros and thermofield dynamics across a quantum critical point},\
  }\href {https://doi.org/10.1103/PhysRevResearch.6.043139} {\bibfield
  {journal} {\bibinfo  {journal} {Phys. Rev. Res.}\ }\textbf {\bibinfo {volume}
  {6}},\ \bibinfo {pages} {043139} (\bibinfo {year}
  {2024}{\natexlab{b}})}\BibitemShut {NoStop}%
\bibitem [{\citenamefont {Meng}\ \emph {et~al.}(2025)\citenamefont {Meng},
  \citenamefont {Lv}, \citenamefont {Liu}, \citenamefont {Tan}, \citenamefont
  {Zhao},\ and\ \citenamefont {Zou}}]{Meng2025arxiv}%
  \BibitemOpen
  \bibfield  {author} {\bibinfo {author} {\bibfnamefont {Y.}~\bibnamefont
  {Meng}}, \bibinfo {author} {\bibfnamefont {S.}~\bibnamefont {Lv}}, \bibinfo
  {author} {\bibfnamefont {Y.}~\bibnamefont {Liu}}, \bibinfo {author}
  {\bibfnamefont {Z.}~\bibnamefont {Tan}}, \bibinfo {author} {\bibfnamefont
  {E.}~\bibnamefont {Zhao}},\ and\ \bibinfo {author} {\bibfnamefont
  {H.}~\bibnamefont {Zou}},\ }\href {https://doi.org/10.48550/ARXIV.2501.09478}
  {\bibinfo {title} {Detecting many-body scars from fisher zeros}} (\bibinfo
  {year} {2025}),\ \Eprint {https://arxiv.org/abs/arXiv:2501.09478}
  {arXiv:2501.09478} \BibitemShut {NoStop}%
\bibitem [{\citenamefont {Liu}\ \emph {et~al.}(2024{\natexlab{c}})\citenamefont
  {Liu}, \citenamefont {Yin},\ and\ \citenamefont {Chen}}]{LiuYinChen2024}%
  \BibitemOpen
  \bibfield  {author} {\bibinfo {author} {\bibfnamefont {J.}~\bibnamefont
  {Liu}}, \bibinfo {author} {\bibfnamefont {S.}~\bibnamefont {Yin}},\ and\
  \bibinfo {author} {\bibfnamefont {L.}~\bibnamefont {Chen}},\ }\bibfield
  {title} {\bibinfo {title} {Imaginary-temperature zeros for quantum phase
  transitions},\ }\href {https://doi.org/10.1103/PhysRevB.110.134313}
  {\bibfield  {journal} {\bibinfo  {journal} {Phys. Rev. B}\ }\textbf {\bibinfo
  {volume} {110}},\ \bibinfo {pages} {134313} (\bibinfo {year}
  {2024}{\natexlab{c}})}\BibitemShut {NoStop}%
\bibitem [{\citenamefont {Yao}\ and\ \citenamefont
  {Wang}(2018)}]{WangZ2018PRL}%
  \BibitemOpen
  \bibfield  {author} {\bibinfo {author} {\bibfnamefont {S.}~\bibnamefont
  {Yao}}\ and\ \bibinfo {author} {\bibfnamefont {Z.}~\bibnamefont {Wang}},\
  }\bibfield  {title} {\bibinfo {title} {Edge states and topological invariants
  of non-hermitian systems},\ }\href
  {https://doi.org/10.1103/PhysRevLett.121.086803} {\bibfield  {journal}
  {\bibinfo  {journal} {Phys. Rev. Lett.}\ }\textbf {\bibinfo {volume} {121}},\
  \bibinfo {pages} {086803} (\bibinfo {year} {2018})}\BibitemShut {NoStop}%
\bibitem [{\citenamefont {Zhu}\ \emph {et~al.}(2014)\citenamefont {Zhu},
  \citenamefont {L\"u},\ and\ \citenamefont {Chen}}]{Chen2014pra}%
  \BibitemOpen
  \bibfield  {author} {\bibinfo {author} {\bibfnamefont {B.}~\bibnamefont
  {Zhu}}, \bibinfo {author} {\bibfnamefont {R.}~\bibnamefont {L\"u}},\ and\
  \bibinfo {author} {\bibfnamefont {S.}~\bibnamefont {Chen}},\ }\bibfield
  {title} {\bibinfo {title} {$\mathcal{PT}$ symmetry in the non-hermitian
  su-schrieffer-heeger model with complex boundary potentials},\ }\href
  {https://doi.org/10.1103/PhysRevA.89.062102} {\bibfield  {journal} {\bibinfo
  {journal} {Phys. Rev. A}\ }\textbf {\bibinfo {volume} {89}},\ \bibinfo
  {pages} {062102} (\bibinfo {year} {2014})}\BibitemShut {NoStop}%
\bibitem [{\citenamefont {Cai}\ \emph {et~al.}(2020)\citenamefont {Cai},
  \citenamefont {Huang},\ and\ \citenamefont {Vincent~Liu}}]{Cai2020CPL}%
  \BibitemOpen
  \bibfield  {author} {\bibinfo {author} {\bibfnamefont {Z.}~\bibnamefont
  {Cai}}, \bibinfo {author} {\bibfnamefont {Y.}~\bibnamefont {Huang}},\ and\
  \bibinfo {author} {\bibfnamefont {W.}~\bibnamefont {Vincent~Liu}},\
  }\bibfield  {title} {\bibinfo {title} {Imaginary time crystal of thermal
  quantum matter},\ }\href {https://doi.org/10.1088/0256-307x/37/5/050503}
  {\bibfield  {journal} {\bibinfo  {journal} {Chinese Physics Letters}\
  }\textbf {\bibinfo {volume} {37}},\ \bibinfo {pages} {050503} (\bibinfo
  {year} {2020})}\BibitemShut {NoStop}%
\bibitem [{\citenamefont {Arouca}\ \emph {et~al.}(2022)\citenamefont {Arouca},
  \citenamefont {Marino},\ and\ \citenamefont {Morais~Smith}}]{Smith2022iTC}%
  \BibitemOpen
  \bibfield  {author} {\bibinfo {author} {\bibfnamefont {R.}~\bibnamefont
  {Arouca}}, \bibinfo {author} {\bibfnamefont {E.~C.}\ \bibnamefont {Marino}},\
  and\ \bibinfo {author} {\bibfnamefont {C.}~\bibnamefont {Morais~Smith}},\
  }\bibfield  {title} {\bibinfo {title} {Non-hermitian quantum gases: a
  platform for imaginary time crystals},\ }\bibfield  {journal} {\bibinfo
  {journal} {Quantum Frontiers}\ }\textbf {\bibinfo {volume} {1}},\ \href
  {https://doi.org/10.1007/s44214-022-00002-0} {10.1007/s44214-022-00002-0}
  (\bibinfo {year} {2022})\BibitemShut {NoStop}%
\bibitem [{\citenamefont {Wakatsuki}\ \emph {et~al.}(2014)\citenamefont
  {Wakatsuki}, \citenamefont {Ezawa}, \citenamefont {Tanaka},\ and\
  \citenamefont {Nagaosa}}]{Wakatsuki2014PRB}%
  \BibitemOpen
  \bibfield  {author} {\bibinfo {author} {\bibfnamefont {R.}~\bibnamefont
  {Wakatsuki}}, \bibinfo {author} {\bibfnamefont {M.}~\bibnamefont {Ezawa}},
  \bibinfo {author} {\bibfnamefont {Y.}~\bibnamefont {Tanaka}},\ and\ \bibinfo
  {author} {\bibfnamefont {N.}~\bibnamefont {Nagaosa}},\ }\bibfield  {title}
  {\bibinfo {title} {Fermion fractionalization to majorana fermions in a
  dimerized kitaev superconductor},\ }\href
  {https://doi.org/10.1103/PhysRevB.90.014505} {\bibfield  {journal} {\bibinfo
  {journal} {Phys. Rev. B}\ }\textbf {\bibinfo {volume} {90}},\ \bibinfo
  {pages} {014505} (\bibinfo {year} {2014})}\BibitemShut {NoStop}%
\bibitem [{\citenamefont {Li}\ \emph {et~al.}(2018)\citenamefont {Li},
  \citenamefont {Zhang}, \citenamefont {Zhang},\ and\ \citenamefont
  {Song}}]{LiSong2018PRB}%
  \BibitemOpen
  \bibfield  {author} {\bibinfo {author} {\bibfnamefont {C.}~\bibnamefont
  {Li}}, \bibinfo {author} {\bibfnamefont {X.~Z.}\ \bibnamefont {Zhang}},
  \bibinfo {author} {\bibfnamefont {G.}~\bibnamefont {Zhang}},\ and\ \bibinfo
  {author} {\bibfnamefont {Z.}~\bibnamefont {Song}},\ }\bibfield  {title}
  {\bibinfo {title} {Topological phases in a kitaev chain with imbalanced
  pairing},\ }\href {https://doi.org/10.1103/PhysRevB.97.115436} {\bibfield
  {journal} {\bibinfo  {journal} {Phys. Rev. B}\ }\textbf {\bibinfo {volume}
  {97}},\ \bibinfo {pages} {115436} (\bibinfo {year} {2018})}\BibitemShut
  {NoStop}%
\bibitem [{\citenamefont {Basu}\ \emph {et~al.}(2022)\citenamefont {Basu},
  \citenamefont {Arovas}, \citenamefont {Gopalakrishnan}, \citenamefont
  {Hooley},\ and\ \citenamefont {Oganesyan}}]{Basu2022PRR}%
  \BibitemOpen
  \bibfield  {author} {\bibinfo {author} {\bibfnamefont {S.}~\bibnamefont
  {Basu}}, \bibinfo {author} {\bibfnamefont {D.~P.}\ \bibnamefont {Arovas}},
  \bibinfo {author} {\bibfnamefont {S.}~\bibnamefont {Gopalakrishnan}},
  \bibinfo {author} {\bibfnamefont {C.~A.}\ \bibnamefont {Hooley}},\ and\
  \bibinfo {author} {\bibfnamefont {V.}~\bibnamefont {Oganesyan}},\ }\bibfield
  {title} {\bibinfo {title} {Fisher zeros and persistent temporal oscillations
  in nonunitary quantum circuits},\ }\href
  {https://doi.org/10.1103/PhysRevResearch.4.013018} {\bibfield  {journal}
  {\bibinfo  {journal} {Phys. Rev. Res.}\ }\textbf {\bibinfo {volume} {4}},\
  \bibinfo {pages} {013018} (\bibinfo {year} {2022})}\BibitemShut {NoStop}%
\bibitem [{\citenamefont {Xie}\ \emph {et~al.}(2024)\citenamefont {Xie},
  \citenamefont {Kolodrubetz}, \citenamefont {Oganesyan},\ and\ \citenamefont
  {Arovas}}]{Xie2024arxiv}%
  \BibitemOpen
  \bibfield  {author} {\bibinfo {author} {\bibfnamefont {W.}~\bibnamefont
  {Xie}}, \bibinfo {author} {\bibfnamefont {M.}~\bibnamefont {Kolodrubetz}},
  \bibinfo {author} {\bibfnamefont {V.}~\bibnamefont {Oganesyan}},\ and\
  \bibinfo {author} {\bibfnamefont {D.~P.}\ \bibnamefont {Arovas}},\ }\href
  {https://doi.org/10.48550/ARXIV.2412.04382} {\bibinfo {title} {Effects of
  non-integrability in a non-hermitian time crystal}} (\bibinfo {year}
  {2024}),\ \Eprint {https://arxiv.org/abs/arXiv:2412.04382} {arXiv:2412.04382}
  \BibitemShut {NoStop}%
\bibitem [{\citenamefont {Lieb}\ \emph {et~al.}(1961)\citenamefont {Lieb},
  \citenamefont {Schultz},\ and\ \citenamefont {Mattis}}]{Lieb1961AOP}%
  \BibitemOpen
  \bibfield  {author} {\bibinfo {author} {\bibfnamefont {E.}~\bibnamefont
  {Lieb}}, \bibinfo {author} {\bibfnamefont {T.}~\bibnamefont {Schultz}},\ and\
  \bibinfo {author} {\bibfnamefont {D.}~\bibnamefont {Mattis}},\ }\bibfield
  {title} {\bibinfo {title} {Two soluble models of an antiferromagnetic
  chain},\ }\href {https://doi.org/10.1016/0003-4916(61)90115-4} {\bibfield
  {journal} {\bibinfo  {journal} {Annals of Physics}\ }\textbf {\bibinfo
  {volume} {16}},\ \bibinfo {pages} {407–466} (\bibinfo {year}
  {1961})}\BibitemShut {NoStop}%
\bibitem [{\citenamefont {Haldane}(1988)}]{Haldane1088prl}%
  \BibitemOpen
  \bibfield  {author} {\bibinfo {author} {\bibfnamefont {F.~D.~M.}\
  \bibnamefont {Haldane}},\ }\bibfield  {title} {\bibinfo {title} {Model for a
  quantum hall effect without landau levels: Condensed-matter realization of
  the "parity anomaly"},\ }\href {https://doi.org/10.1103/PhysRevLett.61.2015}
  {\bibfield  {journal} {\bibinfo  {journal} {Phys. Rev. Lett.}\ }\textbf
  {\bibinfo {volume} {61}},\ \bibinfo {pages} {2015} (\bibinfo {year}
  {1988})}\BibitemShut {NoStop}%
\bibitem [{\citenamefont {Qi}\ \emph {et~al.}(2006)\citenamefont {Qi},
  \citenamefont {Wu},\ and\ \citenamefont {Zhang}}]{QiWuZ2006PRB}%
  \BibitemOpen
  \bibfield  {author} {\bibinfo {author} {\bibfnamefont {X.-L.}\ \bibnamefont
  {Qi}}, \bibinfo {author} {\bibfnamefont {Y.-S.}\ \bibnamefont {Wu}},\ and\
  \bibinfo {author} {\bibfnamefont {S.-C.}\ \bibnamefont {Zhang}},\ }\bibfield
  {title} {\bibinfo {title} {Topological quantization of the spin hall effect
  in two-dimensional paramagnetic semiconductors},\ }\href
  {https://doi.org/10.1103/PhysRevB.74.085308} {\bibfield  {journal} {\bibinfo
  {journal} {Phys. Rev. B}\ }\textbf {\bibinfo {volume} {74}},\ \bibinfo
  {pages} {085308} (\bibinfo {year} {2006})}\BibitemShut {NoStop}%
\bibitem [{\citenamefont {Bernevig}\ \emph {et~al.}(2006)\citenamefont
  {Bernevig}, \citenamefont {Hughes},\ and\ \citenamefont {Zhang}}]{BHZ2006}%
  \BibitemOpen
  \bibfield  {author} {\bibinfo {author} {\bibfnamefont {B.~A.}\ \bibnamefont
  {Bernevig}}, \bibinfo {author} {\bibfnamefont {T.~L.}\ \bibnamefont
  {Hughes}},\ and\ \bibinfo {author} {\bibfnamefont {S.-C.}\ \bibnamefont
  {Zhang}},\ }\bibfield  {title} {\bibinfo {title} {Quantum spin hall effect
  and topological phase transition in hgte quantum wells},\ }\href
  {https://doi.org/10.1126/science.1133734} {\bibfield  {journal} {\bibinfo
  {journal} {Science}\ }\textbf {\bibinfo {volume} {314}},\ \bibinfo {pages}
  {1757–1761} (\bibinfo {year} {2006})}\BibitemShut {NoStop}%
\bibitem [{\citenamefont {Kitaev}\ and\ \citenamefont
  {Preskill}(2006)}]{KitaevPreskill2006TEE}%
  \BibitemOpen
  \bibfield  {author} {\bibinfo {author} {\bibfnamefont {A.}~\bibnamefont
  {Kitaev}}\ and\ \bibinfo {author} {\bibfnamefont {J.}~\bibnamefont
  {Preskill}},\ }\bibfield  {title} {\bibinfo {title} {Topological entanglement
  entropy},\ }\href {https://doi.org/10.1103/PhysRevLett.96.110404} {\bibfield
  {journal} {\bibinfo  {journal} {Phys. Rev. Lett.}\ }\textbf {\bibinfo
  {volume} {96}},\ \bibinfo {pages} {110404} (\bibinfo {year}
  {2006})}\BibitemShut {NoStop}%
\bibitem [{\citenamefont {Levin}\ and\ \citenamefont
  {Wen}(2006)}]{LevinWen2006prl}%
  \BibitemOpen
  \bibfield  {author} {\bibinfo {author} {\bibfnamefont {M.}~\bibnamefont
  {Levin}}\ and\ \bibinfo {author} {\bibfnamefont {X.-G.}\ \bibnamefont
  {Wen}},\ }\bibfield  {title} {\bibinfo {title} {Detecting topological order
  in a ground state wave function},\ }\href
  {https://doi.org/10.1103/PhysRevLett.96.110405} {\bibfield  {journal}
  {\bibinfo  {journal} {Phys. Rev. Lett.}\ }\textbf {\bibinfo {volume} {96}},\
  \bibinfo {pages} {110405} (\bibinfo {year} {2006})}\BibitemShut {NoStop}%
\bibitem [{\citenamefont {Li}\ and\ \citenamefont
  {Haldane}(2008)}]{LiHaldane2008prl}%
  \BibitemOpen
  \bibfield  {author} {\bibinfo {author} {\bibfnamefont {H.}~\bibnamefont
  {Li}}\ and\ \bibinfo {author} {\bibfnamefont {F.~D.~M.}\ \bibnamefont
  {Haldane}},\ }\bibfield  {title} {\bibinfo {title} {Entanglement spectrum as
  a generalization of entanglement entropy: Identification of topological order
  in non-abelian fractional quantum hall effect states},\ }\href
  {https://doi.org/10.1103/PhysRevLett.101.010504} {\bibfield  {journal}
  {\bibinfo  {journal} {Phys. Rev. Lett.}\ }\textbf {\bibinfo {volume} {101}},\
  \bibinfo {pages} {010504} (\bibinfo {year} {2008})}\BibitemShut {NoStop}%
\bibitem [{\citenamefont {Chen}\ \emph {et~al.}(2011)\citenamefont {Chen},
  \citenamefont {Gu},\ and\ \citenamefont {Wen}}]{SPT2011prb1}%
  \BibitemOpen
  \bibfield  {author} {\bibinfo {author} {\bibfnamefont {X.}~\bibnamefont
  {Chen}}, \bibinfo {author} {\bibfnamefont {Z.-C.}\ \bibnamefont {Gu}},\ and\
  \bibinfo {author} {\bibfnamefont {X.-G.}\ \bibnamefont {Wen}},\ }\bibfield
  {title} {\bibinfo {title} {Classification of gapped symmetric phases in
  one-dimensional spin systems},\ }\href
  {https://doi.org/10.1103/PhysRevB.83.035107} {\bibfield  {journal} {\bibinfo
  {journal} {Phys. Rev. B}\ }\textbf {\bibinfo {volume} {83}},\ \bibinfo
  {pages} {035107} (\bibinfo {year} {2011})}\BibitemShut {NoStop}%
\end{thebibliography}
%

\end{document}